\documentclass{article}
\usepackage[latin1]{inputenc}
\usepackage[T1]{fontenc}
\usepackage[english]{babel}
\usepackage{latexsym}
\usepackage{graphicx}
\usepackage{epsfig}
\newtheorem{theorem}{Theorem}
\newtheorem{lemma}{Lemma}
\begin{document}
\title{A reciprocal Wald theorem for varying gravitational function}
\author{Stéphane Fay\\
Laboratoire Univers et Théories, CNRS-FRE 2462\\
Observatoire de Paris, F-92195 Meudon Cedex\\
France\footnote{Steph.Fay@Wanadoo.fr}}
\maketitle
\begin{abstract}
We study when a cosmological constant is a natural issue if it is mimicked by the potential of a massive Hyperextended Scalar Tensor theory with a perfect fluid for Bianchi type I and V models. We then deduce a reciprocal Wald theorem giving the conditions such that the potential tends to a non vanishing constant when the gravitational function varies. We also get the conditions allowing the potentiel to vanish or diverge.
\\
\\
\\
Pacs: 04.50.+h, 98.80.Cq, 11.27.+d
\\
\\
\\
European Physical Journal C copyright 2003\\
Eur. Phys. J. C 30, d01, 007, 2003.\\
http://www.springerlink.com
\end{abstract}
\section{Introduction} \label{s1}
In this work, we look for the potential asymptotical behaviours when we consider a Hyperextended Scalar Tensor theory with a massive scalar field and a perfect fluid for the Bianchi type $I$ and $V$ models .

Recent observations have shown that Universe dynamics is accelerated\cite{Per99,Spe03}. This behaviour brings interest on a new standard model which takes into account a cosmological constant $\Lambda$. Hence, it becomes more and more urgent to find an answer to the cosmological constant problem consisting in the huge discrepancy between the observed value of $\Lambda$ and its much larger values predicted by particle physics models at early times. One way to study it is to consider a scalar tensor theory with a massive scalar field whose potential $U$ mimics the evolution of a dynamical $\Lambda$ and is thus able to tend to a constant for late times. Scalar tensor theories are a prediction of particle physics theories which admit them as low energy limits and, most of time, also imply that the gravitational constant $G$ is a varying quantity. The most famous scalar tensor theory with varying $G$ is the Brans-Dicke theory\cite{BraDic61} and has been studied in the sixties: the gravitational function varies as the inverse of a scalar field $\phi$ and the coupling between $\phi$ and the metric is the Brans-Dicke coupling constant $\omega$. Although it is a very interesting theory it is also a very special one. Most of time, low energy limits of particle physics theories imply different forms for $G$ and $\omega$. For instance some string theories predict that $G$ and $\omega$ should respectively evolve as $e^{-\phi}$ and $\phi e^{-\phi}$. Presently, if we consider that our Universe is correctly described by a scalar tensor theory, nobody knows what should be the exact forms of the three functions $G$, $\omega$ and $U$ with respect to $\phi$. Thus, it seems advantageous to consider them as unspecified functions of the scalar field and to use observations to constraint their asymptotical forms. Generally speaking, the class of theories defined by unknown forms of $G$, $\omega$ and $U$ is called Hyperextended Scalar Tensor theories (HST)\cite{TorVuc96} and can be thought of as a modern extension of Dirac's ideas on time variation of constants of nature\cite{Dir37}. This subject has gained renewed interest since the works of \cite{CarMar02}.\\
What teach us the observations about the asymptotical behaviours of $G$, $\omega$ and $U$? Concerning $G$, variation measurements of the gravitational function show that it should tend to a constant. Hence, Viking lander ranging or pulsar-white dwarf binary give $\dot{G}G^{-1}=10^{-12}yr^{-1}$ (see \cite{Chi01} for references). It means that the scalar field would become minimally coupled at late time or that $G$ is vanishing as implied by dimensional analysis of Dirac or variation of fine structure constant\cite{BerTre01}. For $\omega$, the solar system tests and some theoretical studies of nucleosynthesis for scalar tensor theories\cite{SerAli96} show that it should be larger than 500. Moreover, future observations of neutron stars spiralling into massive black holes\cite{SchWil01} could give values larger than $240000$. For the potential, as written above, it should tend to a cosmological constant, vanishing or not. To determine this last point, more precise measurements on $\Lambda$ variation should be performed.\\

What about the geometrical framework of this paper? The standard model is based on the assumptions that our Universe is perfectly homogeneous and isotropic and thus described by FLRW metrics. BOOMERANG\cite{Lan01} and WMAP\cite{Spe03} observations have shown that this description was already correct during the last scattering period. However, FLRW models are very special ones and their behaviour close to singularities is far less general than the one of other models that do not rest on such symetry. Hence, it seems more reasonable to suppose that Universe has naturally evolved toward isotropy and homogeneity at late times but was initially not so symmetric. One way to generalise FLRW geometry is to keep only the homogeneity assumption. Bianchi models describe such geometry. The most realistic choice would be to consider fully inhomogeneous models. However, they do not have been fully classified. They lack of symetry contrary to the Bianchi models whose approach of singularity could be shared by most of the inhomogeneous models. Bianchi's classification includes nine types. Some of them are particularly interesting since they share the same structure constants as the FLRW models and thus may isotropise toward them. Hence the Bianchi type $I$, $V$ and $IX$ models may respectively tend toward flat, open and closed FLRW models. In this work we will choose to study the two first ones.\\

The goal of this paper will be to look for the conditions such that the potential tends to a cosmological constant when Bianchi type $I$ or $V$ models are considered with a non minimally coupled massive scalar field and a perfect fluid. If this constant is asymptotically vanishing, one possible answer to the cosmological constant problem is that the Universe is old. If it is not vanishing, the cosmological constant problem may persist but if the constant is fine tuned. In any case, the potential must tend to a constant so that an answer to this problem could exist in the HST framework. To reach our goal, we will need to express the field equations solutions as some functions of $G$ and of the isotropic part of the metric, $e^{\Omega}$. Then we will look for conditions on these quantities such that the potential tends to a constant. This paper is organised as follows. In section \ref{s2} we write the HST field equations for the Bianchi type $I$ and $V$ models. In section \ref{s3} we determine their solutions with help of quadratures depending on $G$, $\Omega$ and $\phi$. In section \ref{s4} we establish all the possible asymptotical behaviours for the potential depending on inequalities between $\Omega$ and $G$. We discuss the results in the last section.
\section{Field equations} \label{s2}
For any Bianchi model, the diagonal form of their metric may be written as:
\begin{equation} \label{1}
ds^2=-dt^2+g_{\mu\mu}(\omega^\mu)^2=-dt^2+e^{2\alpha}(\omega^1)^2+e^{2\beta}(\omega^2)^2+e^{2\gamma}(\omega^3)^2
\end{equation}
Here, the $\omega^i$ are the 1-forms specifying the Bianchi type $I$ or $V$ models. For the Bianchi type $I$ model, they are $dx$, $dy$ and $dz$, the structure constants beeing zero. For the Bianchi type $V$ model, we have chosen as basis $\omega_1=dx$, $\omega_2=e^{-x}dy$ and $\omega_3=e^{-x}dz$ from which we deduce that all the structure constants are zero but $C^2_{12}=-C^2_{21}=C^3_{13}=-C^3_{31}=-1$. Following Misner\cite{Mis69}, we will use the decomposition of each metric function into an isotropic and anisotropic part:
\begin{eqnarray*}
\alpha&=&\Omega+\beta_+\\
\beta&=&\Omega+\beta_-\\
\gamma&=&\Omega-\beta_+-\beta_-\\
\end{eqnarray*}
The HST action with a potential and a perfect fluid is written:
\begin{equation} \label{2}
S=\int (G^{-1}R-\omega\phi^{-1}\phi_{,\mu}\phi^{,\mu}-U+16\pi L_m)\sqrt{-g}
\end{equation}
$G$ is the gravitational coupling function, $\omega$ the Brans-Dicke coupling function, $U$ the potential and $L_m$ the Lagrangian density of the perfect fluid. Varying (\ref{2}) with respect to the metric functions and scalar field, we get respectively:
\begin{equation} \label{3}
R_{\mu\nu}-\frac{1}{2}g_{\mu\nu}R=G\left[\frac{\omega}{\phi}\phi_{,\mu}\phi_{,\nu}-\frac{\omega}{2\phi}\phi_{,\lambda}\phi^{\lambda}g_{\mu\nu}+(G^{-1})_{,\mu;\nu}-g_{\mu\nu}\Box (G^{-1})-\frac{1}{2}U g_{\mu\nu}+\frac{8\pi}{c^4}T_{\mu\nu}\right]
\end{equation}
\begin{equation} \label{3cs}
\dot{\phi}^{2}\left[-\frac{\omega_{\phi}}{\phi}+\frac{\omega}{\phi^{2}}-G(G^{-1})_{\phi}\frac{\omega}{\phi}\right]+\frac{2\omega}{\phi}\Box \phi+3G(G^{-1})_{\phi}\Box (G^{-1})+2GG^{-1}_\phi U-U_\phi-\frac{8\pi}{c^4}GG^{-1}_\phi T=0
\end{equation}
where $T_{\mu\nu}=(\rho+p)u_{\mu}u_{\nu}+g_{\mu\nu}p$ is the energy-momentum tensor, $\rho$ the density, $p$ the pressure, $u$ the fluid velocity and the dot means a derivative with respect to $t$. In this work, we will consider a perfect fluid with equation of state $p=(\delta-1)\rho$ and $\delta\in\left[1,2\right]$. For a radiation or matter dominated Universe, we have respectively $\delta=4/3$ or $\delta=1$. Consequently, from the energy impulsion conservation law $T^{0\mu}{;\mu}=0$, we deduce that $\rho=e^{-3\delta\Omega}$. We define the $\tau$ time by $dt=e^{\alpha+\beta+\gamma}d\tau$ and introduce it in (\ref{3}). It yields for the spatial field equations components:
\begin{eqnarray*}
\alpha''G^{-1}+\alpha'(G^{-1})'+(1/2G^{-1})''-2\sigma G^{-1}e^{2\beta+2\gamma}-e^{6\Omega}\left[1/2U+4\pi (2-\delta)\rho_0 e^{-3\delta\Omega}\right]=0&&\label{4a}\\
\beta''G^{-1}+\beta'(G^{-1})'+1/2(G^{-1})''-2\sigma G^{-1}e^{2\beta+2\gamma}-e^{6\Omega}\left[1/2U+4\pi (2-\delta)\rho_0 e^{-3\delta\Omega}\right]=0&&\label{4b}\\
\gamma''G^{-1}+\gamma'(G^{-1})'+1/2(G^{-1})''-2\sigma G^{-1}e^{2\beta+2\gamma}-e^{6\Omega}\left[1/2U+4\pi (2-\delta)\rho_0 e^{-3\delta\Omega}\right]=0&&\label{4c}\\\nonumber
\end{eqnarray*}
A prime means a derivative with respect to $\tau$ and $\sigma$ is equal to 0 or 1 depending on respectively Bianchi type $I$ or $V$ models. The constraint equation is written:
\begin{equation} \label{5}
\alpha'\beta'+\alpha'\gamma'+\beta'\gamma'+3\Omega^,GG^{-1,}+3\sigma e^{2\beta+2\gamma}-1/2Ge^{6\Omega}U-1/2G\omega\phi'^{2}\phi^{-1}
-8\pi\rho_0Ge^{3(2-\delta)\Omega}=0
\end{equation}
The Bianchi type $V$ model has an additional constraint which is:
\begin{equation} \label{5a}
2\alpha'-\beta'-\gamma'=0
\end{equation}
A first integral of the Klein-Gordon equation (\ref{3cs}) is contained in the constraint equation (\ref{5}). In the next section, we will express the field equations solutions depending on $G$, $\Omega$ and $\phi$.
\section{Solution of the field equations as functions of $G$, $\Omega$ and $\phi$} \label{s3} We begin to calculate the anisotropic parts $\beta_\pm$ of the metric. Using the above parametrisation and adding the three spatial field equations components, we get:
\begin{equation} \label{6}
3\Omega''G^{-1}+3\Omega'(G^{-1})'+3/2(G^{-1})''-6\sigma G^{-1}e^{2\beta+2\gamma}-3e^{6\Omega}\left[1/2U+4\pi\rho_0(2-\delta)e^{-3\delta\Omega}\right]=0
\end{equation}
We deduce from this expression that:
\begin{equation}\label{5b}
-\Omega'G^{-1}=\int (1/2(G^{-1})''-2\sigma G^{-1}e^{2\beta+2\gamma}-e^{6\Omega}(1/2U+4\pi\rho_0(2-\delta)e^{-3\delta\Omega}))d\tau
\nonumber
\end{equation}
Taking into account the Bianchi type $V$ additional constraint (\ref{5a}), we get for this model that $\beta_+=\beta_{+1}$ is a constant, showing that anisotropy is mainly described by the function $\beta_-$. In this case, the metric takes the form $ds^2=-dt^2+e^{2\Omega}(\omega^1)^2+e^{2\Omega+2\beta_-}(\omega^2)^2+e^{2\Omega-2\beta_-}(\omega^3)^2$. Introducing expression (\ref{5b}) in the second spatial component, we get after integration:
\begin{equation} \label{7}
\beta_-=\beta_{-0}\int Ge^{-3\Omega}dt + \beta_{-1}
\end{equation}
where $\beta_{-0}$ and $\beta_{\pm 1}$ are integration constants. This expression is valid for both Bianchi type $I$ and $V$ models. For the Bianchi type $I$ model, the $\beta_+$ function takes the same form as (\ref{7}). In \cite{Fay00C}, the same calculi have been done for the Bianchi type $I$ model without a perfect fluid and the same forms for $\beta_\pm$ have been found: hence, the expression of the anisotropic part of the metric is not modified by the presence of a perfect fluid.\\
Now, it is possible to express the potential as a function of $G$ and $\Omega$. From the equation (\ref{6}) and expressions for $\beta_\pm$, we derive for the potential:
\begin{equation} \label{8}
U=2(\ddot{\Omega}+3\dot{\Omega}^2)G^{-1}+5\dot{G^{-1}}\dot{\Omega}+\ddot{G^{-1}}-4\sigma G^{-1}e^{-2\Omega-2\beta_{+1}}-8\pi\rho_0(2-\delta)e^{-3\delta\Omega}
\end{equation}
The overdot means a derivative with respect to proper time $t$. From the constraint equation and the expression (\ref{8}) for the potential, we deduce the form of the Brans-Dicke coupling function:
\begin{equation}\label{9} \omega=2G^{-1}\dot{\phi}^{-2}\phi\left[-\ddot{\Omega}-3\dot{\Omega}^2+1/2G\dot{G^{-1}}\dot{\Omega}-1/2G\ddot{G^{-1}}+5\sigma e^{-2\Omega-2\beta_{+1}}-4\pi\rho_0\delta G e^{-3\delta\Omega}\right]
\end{equation}
The equations (\ref{7}-\ref{9}) constitute a solution of the HST field equations for the Bianchi type $I$ and $V$ models depending on the quantities $\Omega$, $G$ and $\phi$. Similar results have been found recently in \cite{RivPim02} for FLRW models and with $G^{-1}=\phi$. Here, we do not specify the gravitational function form such that it is clear that our results apply to any form of $G(\phi)$. We could also use a conformal transformation and thus only study a minimally coupled scalar tensor theory. However, reversing the conformal transformation to recover some results for varying $G$ is not always workable and thus we discard this method from this paper.
\section{Potential asymptotical behaviour} \label{s4}
In what follows, we are going to determine what are the potential asymptotical dominant terms depending on conditions on $G$ and $\Omega$. To reach this goal, we will use the following lemma:
\begin{lemma}
Let $F_1$ and $F_2$ be two positive functions and let us define the sign "$<$" as meaning "$<<$" or $\propto$. 
For any two positive functions $F_1$ and $F_2$ such that $\mid F_1\mid<\mid F_2\mid$, when $t\rightarrow +\infty$ then $\mid \dot F_1\mid<\mid\dot F_2\mid$. Here, $\mid\mid$ stands for absolute value when $F_i$ is mototonic and for absolute and mean values when it is oscillating.
\end{lemma}
This lemma is always true for monotonic functions but is limited for oscillating ones. Hence, when $F_1$ or/and $F_2$ are oscillating functions such that $\mid F_1\mid<\mid F_2\mid$, writing that the means values of $\dot F_1$ and $\dot F_2$ are such that $\mid \dot F_1\mid<\mid\dot F_2\mid$ means that asymptotically the derivative of $F_1$ have to be continually larger than the one of $F_2$. Hence, the lemma does not hold for $F_1=3\Omega^{-1}$ and $F_2=sin^2(10\Omega)\Omega^{-1}$ but for $F_1=3\Omega^{1/2}$ and $F_2=sin^2(10\Omega)\Omega^{-1}$. When the lemma is true, $\mid F_1 \mid>\mid 1 \mid$ means that $F_1$ is asymptotically larger or proportional to a normalized constant. $\mid F_1 \mid>\mid F_2 \mid$ means that $F_1$ is asymptotically larger or proportional to the function $F_2$: in the first case, the left term is negligible with respect to the right one, in the second case, we can choose arbitrarily which term we want to neglect since our goal is to know the asymptotical behaviour of the potential. \\
For sake of simplicity, we will put $K=G^{-1}$. Following the lemma, we have the following inequalities:
\begin{itemize}
\item $\mid\Omega\mid <\mid \ln t\mid\Rightarrow \mid \dot\Omega\mid <\mid t^{-1}\mid\Rightarrow \mid\dot\Omega\mid^{-1}>\mid t\mid\Rightarrow \mid\frac{\ddot\Omega}{\dot\Omega^2}\mid >\mid 1\mid\Rightarrow \mid\dot\Omega^2 \mid < \mid\ddot\Omega\mid$
\item $\mid\Omega\mid < \mid\ln K\mid \Rightarrow \mid \dot\Omega\mid<\mid \dot K K^{-1}\mid\Rightarrow \mid K \dot\Omega \mid<\mid \dot K\mid \Rightarrow \mid K\dot\Omega^2 \mid<\mid \dot K \dot \Omega\mid$
\item $\mid \Omega\mid<\mid \ln \dot K\mid\Rightarrow \mid \dot\Omega \mid <\mid \ddot K \dot{K}^{-1} \mid \Rightarrow \mid \dot K \dot\Omega \mid <\mid \ddot K \mid$
\item $\mid e^\Omega \mid < \mid K^{\frac{1}{2-3\delta}} \mid \Rightarrow \mid e^{(3\delta-2)\Omega} \mid < \mid K^{-1} \mid \Rightarrow  \mid Ke^{-2\Omega} \mid < \mid e^{-3\delta\Omega} \mid$
\item $\mid \Omega \mid > \mid \ln t \mid \Rightarrow \mid \dot\Omega e^\Omega \mid > \mid 1 \mid \Rightarrow \mid e^{-2\Omega} \mid < \mid \dot\Omega^2 \mid \Rightarrow \mid Ke^{-2\Omega} \mid < \mid K\dot\Omega^2 \mid$\footnote{In this inequality, we have used $\mid \Omega \mid > \mid \ln t \mid\Rightarrow \mid e^\Omega \mid > \mid t \mid$. This is correct if $\Omega$ and $t$ diverge positively , that we will assume in the discussion.}
\item $\mid \ln \dot\Omega \mid < \mid \ln K \mid \Rightarrow \mid \frac{\ddot \Omega}{\dot \Omega} \mid < \mid \frac{\dot K}{K} \mid \Rightarrow \mid K\ddot\Omega \mid < \mid \dot K \dot\Omega \mid$\footnote{Note that, in general, $\mid F_1 \mid < \mid F_2 \mid\not \Rightarrow \mid \ln F_1 \mid < \mid \ln F_2 \mid$ because $\mid \ln \mid$ is not a monotonic function and, for instance, $F_1$ may vanishes faster than $F_2$ diverges.}
\end{itemize}
From them we get the figure 1, showing the potential asymptotical dominating terms at late times. Note that an identical figure exists for the quantity $\omega\dot{\phi}^{2}\phi^{-1}$. In the next section, from figure 1, we discuss in which conditions the potential may tend to a constant, vanishing or not,  for late times.
\section{Discussion} \label{s5}
In this discussion, we will assume that the Universe is expanding at late times, i.e. $\Omega\rightarrow +\infty$ when $t\rightarrow +\infty$ and we will look for the potential asymptotical behaviour.\\
\\
We first consider the case for which $K$ tends to a non vanishing constant. Then, in figure 1, the only branches we have to examine are such that $\mid \Omega \mid >\mid\ln K \mid$. Moreover, we will have $\ddot K\rightarrow 0$ and $e^{-3\delta\Omega}<Ke^{-2\Omega}$. This last inequality means that \emph{the curvature will not change the general behaviour of the potential when it will diverge or tend to a non vanishing constant. It may only affect the way it vanishes.}\\
Hence, \underline{in presence of a late time accelerated expansion} for the Bianchi type $I$ model, when $\mid \Omega \mid < \mid\ln \dot{K}\mid$, the potential tends to the dominating term among $(\dot{\Omega}^2K, \ddot K, e^{-3\delta\Omega})$. The two last ones asymptotically vanish. It diverges as $\dot\Omega^2$ if $\Omega >> t$. It will tend to a constant if $\Omega \propto t$. Otherwise, it will be vanishing. It is the same if $\mid \Omega \mid > \mid \ln \dot{K}\mid$ or for the Bianchi type $V$ model. As above written, the only differences come in the case for which $U$ vanishes: the variation of $U$ to $0$ may differ depending on which model we consider.\\
For the Bianchi type $I$ model, \underline{when no late time accelerated expansion is present} and $\mid \Omega \mid < \mid\ln \dot{K}\mid$, the potential tends to the dominating term among $(\ddot\Omega K, \ddot K, e^{-3\delta\Omega})$. The two last ones are always asymptotically vanishing. Since $\mid \ddot \Omega \mid < t^{-2}$, the first one vanishes too. Hence, the only possibility is that the potential vanishes. It is the same if we consider that $\mid \Omega \mid > \mid\ln \dot{K}\mid$ or the Bianchi type $V$ model. However, once again, the variation of the potential to $0$ may differ since, for instance, for the Bianchi type $V$ model, $e^{-3\delta\Omega}$ have to be replaced by $K e^{-2\Omega}$ or, when $\mid \Omega \mid > \mid\ln \dot{K}\mid$, the $\ddot K$ term disappears.\\
Thus we get the following result allowing to know if a cosmological constant is naturally explained by HST when the gravitational function tends to a constant:
\begin{theorem}
When the Universe is asymptotically expanding and the gravitational function tends to a non vanishing constant, the potential tends to a cosmological constant only if $e^\Omega\rightarrow e^{\alpha t}$, $\alpha$ being a constant. The potential vanishes when $e^{\Omega}$ varies slower than $e^{\alpha t}$ and diverges otherwise.
\end{theorem}
An important case from the point of view of isotropisation is when $e^\Omega\rightarrow t^m$. Then, we calculate that $U$ vanishes as $t^{-2}$ for the Bianchi type $I$ model and as $t^{-2}$ or $t^{-2m}$ when respectively a late time accelerated expansion occurs or not for the Bianchi type $V$ model. These results are in agreement with isotropisation of the Bianchi type $I$ and $V$ models studied in \cite{Fay01} and \cite{prep}. For the Bianchi type $I$ model, it has been shown that when the scalar field is minimally coupled, the metric functions tend toward a power or exponential laws of time and the potential decreases as $t^{-2}$ or a constant respectively. For the Bianchi type $V$ model, the same result holds since isotropisation always leads to late time accelerated expansion and flat Universe.\\
\\
As a second case, we assume that $K$ diverges slower than or in the same way as $t^2$. Then $1<<\mid K\mid < t^2$, thus implying $\mid \dot K\mid <\mid t\mid$ and $\mid \ddot K\mid<1$.\\
\underline{When a late time accelerated expansion arises}, we will always have $\mid\Omega\mid > \mid\ln \dot K\mid$ since the maximum value of $\dot{K}$ is $t$. Hence we will only consider these branches on figure 1.\\
If $\mid\Omega\mid < \mid\ln K \mid$, the potential will tend to a constant if it is the same for $\dot \Omega \dot K$. Since we have $\mid\Omega\mid <\mid\ln K\mid$, it follows that $\mid\dot \Omega \dot K\mid <\mid\dot K^2 K^{-1}\mid$. Thus, when $U\rightarrow \Lambda$, this last inequality is such that $1<\mid\dot K^2 K^{-1}\mid$ which implies that $\mid K\mid > t^2$  whereas we have assumed it is smaller than $t^2$. It follows that the potential will tend to a non vanishing constant only if $\mid K\mid\propto t^2$ and $\mid \Omega\mid \propto\mid\ln t\mid$. Otherwise, it may diverge or vanish if respectively $\dot{\Omega}\dot{K}$ diverges or vanishes.\\
If $\mid\Omega\mid >\mid\ln K\mid$, $U\rightarrow \Lambda$ only if it is the same for $\dot \Omega^2K$. Except the above inequalities defining this case, there is no additionnal restriction on the asymptotical behaviours for $K$ and $\Omega$ allowing this behaviour. For instance, $U$ is a constant when $K\rightarrow t^{3/2}$ and $\dot{\Omega}\rightarrow t^{-3/4}$: one easily checks that the expansion is accelerated, $\mid\Omega\mid >\mid\ln K\mid$ and $\mid\Omega\mid >\mid\ln \dot K \mid$. Hence the asymptotical behaviour for the potential is ruled by the product $\dot \Omega^2K$: $U$ may tend to a constant, diverge or vanish if respectively $\dot{\Omega}^2 K$ tends to a constant, diverges or vanishes.\\
\underline{When there is no late time accelerated expansion}, for the Bianchi type $I$ model with $\mid \Omega\mid <\mid\ln \dot{K}\mid$ and $\mid\ln \dot \Omega\mid <\mid\ln K \mid$, the potential will tend to a constant if $K\propto t^2$ only. For the other cases, let us assume that $K$ is smaller than $t^2$. Then, the potential will tend to a constant only if it is the case for $Ke^{-2\Omega}$ or $K\ddot \Omega$ or $\dot\Omega\dot K$. However, it will occurs if $e^{-2\Omega}$ or $\ddot\Omega$ or $\dot\Omega$ are larger than respectively $t^{-2}$ or $t^{-2}$ or $t^{-1}$. But this is impossible since there is no accelerated expansion. Hence, $\mid K\mid\propto t^2$ and $\mid\Omega\mid\propto \mid\ln t\mid$ are necessary and sufficient conditions such that the potential tends to a constant.\\
If the potential does not become a constant, it vanishes for the Bianchi type $I$ model since $\mid\Omega\mid < \mid\ln t\mid$ and $\mid K\mid <t^2$. It diverges or vanishes for the Bianchi type $V$ model respectively depending on the behaviour of $Ke^{-2\Omega}$.\\
\\
A third case is when $\mid K\mid>> t^2$ and thus $\mid\ddot K \mid>>1$. Then $K$, $\dot K$ and $\ddot K$ have the same sign and the potential will diverge but if it exists one term able to cancel the divergence of $\ddot K$ at least. Let us examine the expression (\ref{8}). It can not be the $e^{-3\delta\Omega}$ term since it vanishes nor the terms containing $\dot\Omega$ since we have assumed an expanding Universe and thus they have the same sign as $\ddot K$. It can not be the $G^{-1}e^{-2\Omega-2\beta_{+1}}$ term because $\beta_{+1}$ is an integration constant and then would cancel $\ddot K$ for some zero measure values of $\beta_{+1}$. Hence, the only terms able to cancel the divergence of $\ddot K$ would be $2\ddot\Omega K$ and should be such that $\ddot\Omega$ be negative (i.e. no late time accelerated expansion). However, in this case, $G^{-1}e^{-2\Omega-2\beta_{+1}}$ diverge because $e^\Omega <t$. Thus, once again, the potential will diverge but for zero measure values of $\beta_{+1}$. Hence, we conclude that, except for zero measure cases of the integration constant $\beta_{+1}$, the potential diverges when $\mid K\mid>> t^2$.\\
\\
The last possiblity is when $K$ vanishes. When there is no late time accelerated expansion, the potential vanishes since the derivatives of $K$ and $\Omega$ also vanish. If there is a late time accelerated expansion and $e^{\Omega}<e^t$, the potential vanishes since $\mid \dot\Omega \mid < 1$. If there is a late time accelerated expansion with $\mid\Omega\mid<\mid\ln K\mid$ and $\mid\Omega\mid<\mid\ln \dot K\mid$, it vanishes again. If $\mid\Omega\mid<\mid\ln K\mid$ and $\mid\Omega\mid>\mid\ln\dot K\mid$, it vanishes or diverges if $\dot\Omega\dot K$ respectively vanishes or diverges. If $\mid\Omega\mid>\mid \ln K\mid$, it vanishes or diverges if $\dot\Omega^2 K$ respectively vanishes or diverges.\\
\\
Again, these results allow knowing if a cosmological constant is naturally explained by HST when the gravitational function does not tend to a constant:
\begin{theorem}
Let $\alpha$ be a constant, $H=\dot\Omega$ the Hubble function and $G$ the gravitational function. When the Universe is asymptotically expanding and $G$ does not tend to a non vanishing constant, we have the following results:
\begin{itemize}
\item When the gravitational function vanishes faster than $t^{-2}$, the potential diverges.
\item When the gravitational function vanishes slower than or as $t^{-2}$, then in presence of a late time accelerated expansion when $\mid\Omega\mid < \mid\ln G^{-1}\mid$, the potential tends to a constant if $\mid G^{-1}\mid\propto t^2$ and $e^\Omega\propto t^\alpha$. Otherwise, it vanishes/diverges if it is the case for $H \dot{G^{-1}}$. When $\mid\Omega\mid > \mid\ln G^{-1}\mid$, the potential tends to a constant, vanishes or diverges if it is the case for $H^2 G^{-1}$.
\item When the gravitational function vanishes slower than or as $t^{-2}$ and there is no late time accelerated expansion, the potential will tend to a constant for the Bianchi type $I$ model if $\mid \Omega\mid <\mid\ln \dot G^{-1}\mid$, $\mid\ln H\mid <\mid\ln G^{-1} \mid$ and $G^{-1}\propto t^2$, for any other cases if $G^{-1}\propto t^2$ and $e^\Omega\propto t^\alpha$. Otherwise it vanishes for the Bianchi type $I$ model, vanishes/diverges for the Bianchi type $V$ model depending on the curvature term $G^{-1}e^{-2\Omega}$.
\item When the gravitational function diverges and the Universe expands slower than $e^{\alpha t}$ or if there is a late time accelerated expansion with $\mid\Omega\mid<\mid\ln G^{-1}\mid$ and $\mid\Omega\mid<\mid\ln \dot{G^{-1}}\mid$, the potential vanishes. If Universe expands faster than $e^{\alpha t}$ and $\mid\Omega\mid>\mid\ln\dot{G^{-1}}\mid$ or $\mid\Omega\mid>\mid\ln G^{-1}\mid$, it vanishes/diverges if respectively $H\dot{G^{-1}}$ or $H^2 (G^{-1})$ vanishes/diverges.
\end{itemize}
\end{theorem}
Wald has shown that for all initially expanding Bianchi Universes (except the Bianchi type IX model) and for General Relativity with a cosmological constant, the Universe asymptotically reaches a De Sitter one\cite{Wal83}. Our results may be interpreted as a \emph{reciprocal Wald theorem extended to the case of a varying $G$}:
\begin{theorem}
Either of the three folowing conditions is sufficient for the cosmological constant asymptotic behaviour:
\begin{itemize}
\item the gravitational function tends to a constant and the isotropic part of the metric toward an exponential law.
\item the gravitational function vanishes as $t^{-2}$ and, generally, the isotropic part of the metric tends toward a power law.
\item the gravitational function vanishes slower than or as $t^{-2}$, the Universe undergoes a late time accelerated expansion with $\mid\Omega\mid > \mid\ln G^{-1}\mid$ and $H^2 G^{-1}$ tends to a constant.
\end{itemize}
\end{theorem}
To illustrate the above results, consider the following asymptotical forms for $e^{\Omega}$ and $K$:
\begin{eqnarray}
e^{\Omega}&=&t^m\\
K&=&K_0+K_1t^n\\
\end{eqnarray}
where $m>0$ and $n$ are some constants. The solution thus defined is the outcome of numerous theories such as the one defined by $G^{-1}=\phi$, $\omega=\omega_0$ and $V=\phi^\alpha$ when Universe isotropises\cite{BilColIba99}. In \cite{ColIbaHoo97}, power law attractors have been found for $e^\Omega$ when $G^{-1}=K_0$ and $U=e^{k\phi}$. We deduce for the potential:
\begin{eqnarray*}\label{}
U &=& -4\sigma e^{-2\beta_{+1}}(K_0+K_1t^n)t^{-2m}+2mK_0(3m-1)t^{-2}+K_1(-2m+6m^2-n\\
&&+5mn+n^2)t^{n-2}+8\pi\rho_0(\delta-2)t^{-3m\delta}\\
\end{eqnarray*}
When $K_1=0$, the gravitational function is a constant and the potential tends to vanish. When $n>2$, $G^{-1}$ vanishes faster than $t^{-2}$ and the potential diverges as $t^{n-2}+t^{n-2m}$. When $K_0=0$ and $0<n\leq 2$, $G^{-1}$ vanishes slower or as $t^{-2}$ and the potential behaves as $\sigma t^{n-2m}+t^{-2}+t^{n-2}+t^{-3m\delta}$. In presence of an accelerated expansion, $m>1$, $t^{n-2m}<t^{n-2}$ and the potential will tend to a constant only if $n=2$. Otherwise, whatever the sign of $\mid\Omega\mid-\mid\ln \dot K\mid$, it will vanish since $n<2$. In the absence of a late time accelerated expansion, $m<1$ and $t^{n-2m}>t^{n-2}$. Once again, the potential will tend to a constant if $n=2$. Otherwise, it vanishes for the Bianchi type $I$ model and vanishes or diverges for the Bianchi type $V$ model if respectively $n-2m$ is negative or positive. Whatever $K_0$, when $n<0$, the potential vanishes. These results are in agreement with what we have written above\\
\\
In this work we have expressed the HST field equations solutions for Bianchi type $I$ and $V$ models as some functions of observational quantities $G^{-1}$, $e^\Omega$ and their derivatives. Then, we have found the inequalities between these quantities ruling the potential asymptotical behaviour and defining some conditions such that it asymptotically mimics a cosmological constant, vanishing or not. We conclude by comparing these conditions with the observations. Since they seem to show a late time accelerated expansion, we find 2 alternatives so that the HST be able to solve the cosmological constant problem. The first alternative is when the gravitational constant tends to a non vanishing one. The above results show that the dynamical behaviour of the metric should be such that $t<e^{\Omega}\leq e^{\alpha t}$. Then the potential tends to a non vanishing constant if the Universe tends toward a De Sitter model, otherwise it vanishes. A second alternative is when $G$ tends to zero. Then, its variation should be like or slower than $t^{-2}$ otherwise the potential would diverge, which will not be compatible with a cosmological constant. The behaviour of the potential depends on the variation of $e^{\Omega}$ with respect to $G^{-1}$. If $e^{\Omega}<G^{-1}$($e^{\Omega}>G^{-1}$), the potential will tend to a non vanishing cosmological constant only if $G^{-1}\rightarrow t^2$ and $e^{\Omega}\rightarrow t^\alpha$(respectively $H^2G^{-1}\rightarrow const$). Otherwise it will vanish if $H\dot{G^{-1}}\rightarrow 0$(respectively $H^2G^{-1}\rightarrow 0$). In both cases, when the potential tends to a non vanising constant thus respecting the reciprocal Wald theorem, the cosmological problem is solved only if the corresponding solution is an attractor, thus avoiding fine tuning problem. Otherwise, when it disappears, the cosmological problem is solved if the Universe is sufficiently old. This work presents new theoritical constraints on observables $e^{\Omega}$, $H$ and $G^{-1}$ allowing to get asymptotically a small cosmological constant compatible with a late time accelerated expansion and a small gravitationnal constant for scalar tensor theories. Checking observationnaly the above limits would qualify or disqualify the presence of a massive scalar field mimicking an effective cosmological constant in our Universe if geometry is of Bianchi types $I$ or $V$. A next step would be to study HST properties with such constraints.

\newpage
\begin{figure}[!t]
\includegraphics[width=\textwidth,height=16.5cm,bb=20 36 534 800,clip=]{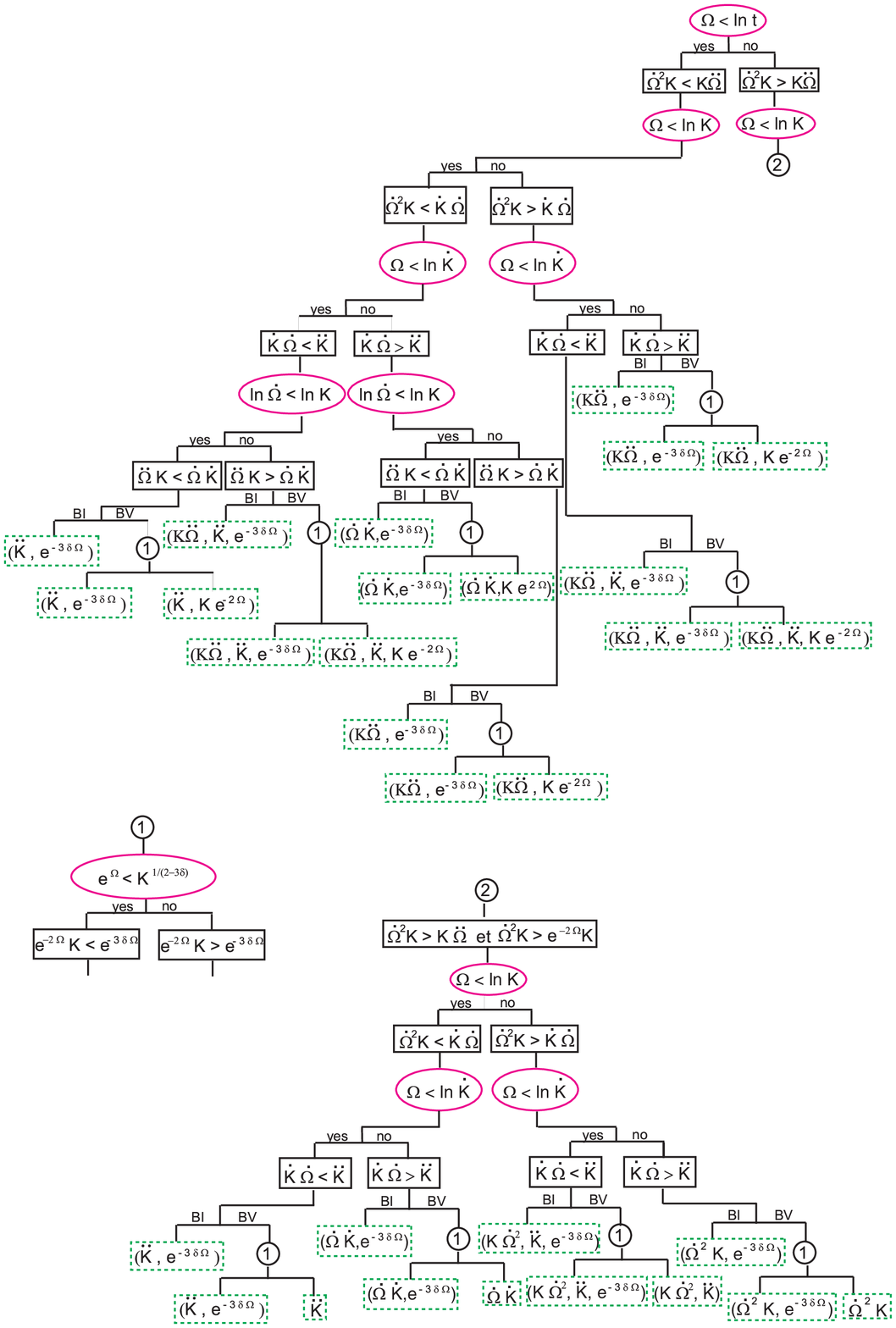}
\caption{\label{fig1} This diagram shows the asymptotical dominant terms in the potential when $t\rightarrow +\infty$ and the Universe expands following the inequalities of section \ref{s4}. Ellipsoid indicates if an equality is (yes) or is not respected (no). Depending on the answer, a quantity can be neglected: this is the meaning of the rectangles that come below each ellipsoid. Dotted rectangles are the dominant term in the potential when all these quantities have been neglected. The number in the circles indicates how two parts of the figure may be joined and $BI$ and $BV$ stand respectively for the Bianchi type $I$ or $V$ model. To clarify the diagram the absolute value symbol "$\mid \mid$" have been omitted.}
\end{figure}
\end{document}